\RequirePackage[2018-12-01]{latexrelease}
\documentclass{rQUF2e}

\usepackage{graphicx}
\usepackage{epstopdf}
\usepackage{subfigure}

\usepackage{amsmath}
\usepackage{amsthm}
\usepackage{url}

\usepackage{booktabs}

\theoremstyle{plain}
\newtheorem{theorem}{Theorem}[section]

\newtheorem{proposition}[theorem]{Proposition}

\theoremstyle{definition}

\theoremstyle{remark}

\begin{document}

\raggedbottom


\title{Learning Markovian Lifts for Rough Heston:\\ Pricing Accuracy with Fewer States}

\author{Arthur Chagas$^{\ast}$$\dag$\thanks{$^\ast$Corresponding author.
Email: arthurchagas@dcc.ufmg.br}, Arthur Buzelin$\dag$, Pedro Bento$\dag$, Yan Aquino$\dag$, Wagner Meira Jr.$\dag$ and Cristiano Arbex Valle$\dag$\\
\affil{$\dag$Computer Science Department, Universidade Federal de Minas Gerais (UFMG), Brazil}}

\maketitle

\begin{abstract}
Markovian lifts make rough-volatility models computationally tractable by approximating fractional memory with a finite sum of exponentials. However, minimising kernel approximation error does not necessarily minimise option pricing error. We propose task-aligned Markovian lifts for rough Heston, learning exponential weights and decay rates directly from vega-normalised pricing errors while retaining the original characteristic-function pricer. A first-order sensitivity analysis explains why pricing accuracy depends on a low-dimensional subspace of kernel perturbations. Across 200 synthetic configurations, including 50 stress scenarios, the proposed lift achieves median paired error ratios of up to 26 relative to BL2 and reduces the median number of states required to reach an implied-volatility error of 0.1 basis points from four to two under standard conditions and from six to four under stress. An amortised construction conditioned on roughness and maturity preserves most of these gains without online optimisation. On 646 weekly SPX option surfaces from 2013--2025, the four-state amortised lift achieves worst-strike implied-volatility errors within one basis point on 98\% of surfaces, compared with 49\% for BL2, and accelerates median calibration fourfold while reproducing the reference calibration fit within one basis point on 94\% of surfaces. 
\end{abstract}


\begin{classcode}
G13; C63; C61; C58
\end{classcode}

\section{Introduction}
\label{sec:introduction}
\label{sec:related}

Rough-volatility models explain important features of financial markets, including the roughness of volatility paths and the steep short-maturity behaviour of implied-volatility surfaces~\citep{gatheral2018volatilityrough,bayer2016pricingrough}. Rough Heston is particularly attractive because it retains the affine structure of classical Heston, allowing European options to be priced through a characteristic function and a fractional Riccati--Volterra equation~\citep{heston1993,eleuch}. However, the fractional memory makes the model non-Markovian and considerably more expensive to use.

Several approaches have been proposed to reduce this cost, including hybrid simulation schemes~\citep{bennedsen2017hybrid} and Monte Carlo variance reduction~\citep{mccrickerd2018turbocharging}. Another approach approximates the fractional kernel by a sum of exponentials, turning the model into a finite-dimensional Markovian system~\citep{abijaber,cuchiero2019markovian}. Each exponential adds one state, so the goal is to obtain accurate prices with as few states as possible. Existing methods mainly rely on quadrature rules and kernel approximation errors. A particularly strong example is the BL2 rule of \citet{bayer2023weakmarkovianapproximationsrough}, designed for European options and supported by weak-error bounds.

Machine learning offers another way to accelerate pricing and calibration. Neural networks have been used to approximate prices and implied-volatility surfaces~\citep{horvath2021deeplearningvolatility,bayer2019deepcalibration,ferguson2018deeply}, while other methods learn solutions to pricing equations or more general solution operators~\citep{sirignano2018dgm,raissi2019physics,lu2019deeponet,li2021fourier}. These approaches typically learn the output of a numerical solver. Here, we instead learn part of the model representation used inside the solver.

Our starting point is a simple question: does a better kernel approximation necessarily give better option prices? A kernel norm measures errors across the entire kernel, but not all of these errors matter equally for pricing. We therefore optimise the weights and decay rates of the exponential modes directly against implied-volatility errors over a family of rough-Heston parameters. The resulting \textbf{task-aligned lift} depends only on the roughness $H$ and maturity $T$, just like a conventional quadrature rule. We consider both positive and signed weights and retain the original model dynamics and characteristic-function pricer. By training the lifts once on a grid over $(H,\log T)$ and interpolating them, we also obtain an amortised version that requires no optimisation at pricing time.

A first-order analysis helps explain why this approach works. The pricing error can be viewed as a quadrature error involving the Laplace transforms of the pricing sensitivities. While kernel-based rules control a much broader family of functions, the task-aligned lift focuses on those that matter for pricing. When these sensitivities span a low-dimensional space, a Tchakaloff-type argument~\citep{bayer2006tchakaloff} shows why a small number of positive states can be sufficient to control the first-order pricing error.

Against BL2 and other kernel-based rules, task-aligned lifts achieve substantially lower pricing errors at matched state budgets, including under stress. These gains persist under amortisation and in an empirical study of weekly SPX surfaces from 2013 to 2025.

Our contributions are threefold: (i) a task-aligned construction of Markovian lifts with an amortised counterpart; (ii) a sensitivity-based quadrature interpretation connecting pricing accuracy to state efficiency; and (iii) synthetic and empirical evidence of improved pricing accuracy, robustness and calibration efficiency.

\section{Method}
\label{sec:method}

\subsection{Rough volatility and its Markovian lift}

Rough stochastic volatility writes the variance as a stochastic Volterra equation driven by the fractional Riemann--Liouville kernel
\begin{equation}
\begin{aligned}
V_t &= V_0 + \int_0^t K_H(t-s)\,b(V_s)\,ds + \int_0^t K_H(t-s)\,\sigma(V_s)\,dW_s, \\
K_H(\tau) &= \frac{\tau^{H-1/2}}{\Gamma(H+\tfrac12)}, \qquad 0<H<\tfrac12 .
\end{aligned}
\label{eq:volterra}
\end{equation}

The power-law kernel makes $V$ non-Markovian and prevents direct use of a low-dimensional Markovian pricing state. A standard approximation replaces $K_H$ by a positive sum of exponentials,
\begin{equation}
K_H(\tau)\approx K^{(m)}(\tau)=\sum_{j=1}^m w_j e^{-x_j\tau}, \qquad w_j,x_j>0,
\label{eq:soe}
\end{equation}
which yields a finite-dimensional Markovian lift. Each exponential mode carries both the drift and diffusion convolutions:
\begin{equation}
\begin{aligned}
Z_t^{(j)}
= \int_0^t e^{-x_j(t-s)}\big[b(V_s^{(m)})\,ds+\sigma(V_s^{(m)})\,dW_s\big], \quad
dZ_t^{(j)}
= \big[-x_jZ_t^{(j)}+b(V_t^{(m)})\big]dt+\sigma(V_t^{(m)})\,dW_t,
\end{aligned}
\label{eq:lift_state}
\end{equation}
and the lifted variance is $V_t^{(m)}=V_0+\sum_{j=1}^m w_jZ_t^{(j)}$. The drift contribution inside the auxiliary state is essential: omitting it would approximate only part of the Volterra equation in Eq.~\eqref{eq:volterra}. The central methodological question is therefore how to choose the positive weights and decay rates $\{w_j,x_j\}_{j=1}^m$ so that the lifted model prices accurately with as few states as possible.

We distinguish between two classes of exponential approximations:
\begin{equation}
    \mathcal{K}_m^{+}
    =
    \left\{
        \sum_{j=1}^{m} w_j e^{-x_j t}
        : w_j>0,\; x_j>0
    \right\},
\quad \text{and} \quad
    \mathcal{K}_m^{\pm}
    =
    \left\{
        \sum_{j=1}^{m} w_j e^{-x_j t}
        : w_j\in\mathbb{R},\; x_j>0
    \right\}.
\end{equation}
The former preserves complete monotonicity and admits the
standard stochastic Markovian interpretation under the usual
admissibility conditions. The latter enlarges the approximation
space but does not automatically inherit these guarantees.
Our central question is how to select the exponential modes
to achieve accurate European option prices with as few
states as possible.

\subsection{A unified characteristic-function pricer}
\label{sec:pricer}

For rough Heston, with $b(v)=\kappa(\theta-v)$, $\sigma(v)=\nu\sqrt v$, and $d\langle W,B\rangle=\rho\,dt$, European options can be priced through a characteristic function driven by a Riccati--Volterra equation. Here, $\kappa$ is the mean-reversion rate, $\theta$ is the long-run variance, $\nu$ is the volatility of variance, $\rho$ is the return--variance correlation, and $V_0$ is the initial variance. For Fourier frequency $u\in\mathbb{R}$, define $a=iu$ and
\(
F(u,x)=\frac12(a^2-a)+(\rho\nu a-\kappa)x+\frac{\nu^2}{2}x^2 .
\label{eq:F}
\)
For any admissible kernel $K$, solve $h=K*F(u,h)$ and set
\begin{equation}
\log\varphi(u,T)=\kappa\theta\int_0^T h(s)\,ds+V_0\int_0^T F(u,h(s))\,ds .
\label{eq:cf}
\end{equation}

With $K=K_H$, the fractional Riccati equation is solved by a predictor--corrector fractional-Adams scheme, providing the reference pricer. With $K=K^{(m)}$, the same formulation becomes a finite $m$-factor Riccati system,
\begin{equation}
\psi_j'(t)=-x_j\psi_j(t)+F(u,\bar\psi(t)), \qquad \bar\psi(t)=\sum_{j=1}^m w_j\psi_j(t),
\label{eq:lifted_riccati}
\end{equation}
with $\psi_j(0)=0$, integrated by an exponential time-differencing scheme to handle stiff decay rates. Both pricers use Fourier cosine series COS inversion for out-of-the-money (OTM) options (puts below the forward and calls above), isolating the kernel approximation error up to numerical discretisation errors. 

The finite-dimensional Riccati representation remains algebraically valid for signed exponential weights. However, stochastic admissibility and the existence of a corresponding non-negative variance process are not guaranteed by this representation alone. Throughout this work, signed approximations are therefore interpreted as numerical surrogates of the rough-Heston Riccati--Volterra equation for European option pricing.

Agreement with closed-form Heston at $H=\tfrac12$ does not establish reference accuracy in the rough regime. We therefore cross-check the reference using an independent solver based on the representation
\(
K_H(\tau)=c_H\int_{\mathbb{R}}e^{(1/2-H)s}\exp(-e^s\tau)\,ds,
\label{eq:reference_kernel}
\)
where $c_H=1/[\Gamma(H+\tfrac12)\Gamma(\tfrac12-H)]$. We discretise the integral by the midpoint rule in $s$, adding a node at $x=0$ to carry the truncated lower tail. The resulting dense SOE ($\approx130$ modes) is priced on a graded time grid. As the two engines share no time-stepping code, their discrepancy provides an independent assessment of reference uncertainty (Sec.~\ref{sec:experiments}).

\subsection{Learned and amortised lifts}
\label{sec:tal}

Instead of fixing the exponential modes by quadrature, we learn them for the pricing task. We consider both positive and signed exponential approximations. For the positive construction, we parameterise $w_j=\exp(q_j)$ and $x_j=\exp(s_j)$. For the signed construction, weights are allowed to take real values, while decay rates remain strictly positive. Both constructions minimise the same pricing objective, differing only in their admissible parameter spaces.

\paragraph{Kernel-optimal initialisation}

The squared $L^2$ kernel error on $[0,T]$ has the closed form
\begin{equation}
\begin{aligned}
\big\|K_H-K^{(m)}\big\|^2_{L^2(0,T)}
={}& \frac{T^{2H}}{2H\,\Gamma(H+\tfrac12)^2}
-2\sum_{j=1}^m w_j x_j^{-(H+1/2)}
P\!\big(H+\tfrac12,x_jT\big)
+\sum_{i,j=1}^m w_iw_j
\frac{1-e^{-(x_i+x_j)T}}{x_i+x_j},
\end{aligned}
\label{eq:l2exact}
\end{equation}
where $P$ is the regularised lower incomplete gamma function \citep{bayer}. For fixed rates, Eq.~\eqref{eq:l2exact} is a convex quadratic in the weights. We compute the optimal non-negative weights by non-negative least squares and optimise only the $m$ log-rates using variable projection, with multiple starts and continuation in $m$. We call this the $L^2_+$ rule. We also construct BL2$_+$ by retaining the nodes of the BL2 rule of \citet{bayer2023weakmarkovianapproximationsrough} and recomputing its weights under the same non-negativity constraint. Both rules serve as baselines and initialisations.

\paragraph{Task-aligned lift}

Let $\eta=(\kappa,\theta,\nu,\rho,V_0)$ denote the remaining rough-Heston parameters, $\Pi$ a distribution over $\eta$, and $\mathcal{G}_{\mathrm{fit}}$ a set of strikes. For a given $(H,T)$, the task-aligned lift minimises the expected squared vega-normalised price error
\begin{equation}
\mathcal{L}_{H,T}(q,s)
=\mathbb{E}_{\eta\sim\Pi}
\frac{1}{|\mathcal{G}_{\mathrm{fit}}|}
\sum_{k\in\mathcal{G}_{\mathrm{fit}}}
\left[
\frac{P_k^{(m)}(\eta;H,T)-P_k^{\mathrm{ref}}(\eta;H,T)}
{\mathcal{V}_k^{\mathrm{ref}}(\eta;H,T)}
\right]^2,
\label{eq:tal}
\end{equation}
where $\mathcal{V}_k^{\mathrm{ref}}$ is the Black--Scholes vega at the reference implied volatility. Each summand approximates the squared implied-volatility error to first order. The expectation is estimated using $n_{\mathrm{tr}}$ independent draws of $\eta$, with reference prices computed once. The resulting lift depends only on $(H,T)$, using the same instance-specific information as quadrature rules.

We solve Eq.~\eqref{eq:tal} by Levenberg--Marquardt in $2m$ unknowns, initialising from BL2$_+$ and $L^2_+$. The Jacobian is computed by forward differences, evaluating all $2m$ perturbations in a batched Riccati solve, with independent $(H,T)$ problems solved in parallel. We select the iterate with the smallest loss on $n_{\mathrm{val}}$ additional draws of $\eta$ and a disjoint strike grid $\mathcal{G}_{\mathrm{val}}$. Retaining both initialisations as candidates ensures that the selected lift is no worse than either under the validation objective. Final evaluation uses held-out configurations and a third, disjoint strike grid.

Replacing $\Pi$ by a point mass gives the per-instance task-aligned lift. Since it requires reference prices of the configuration being priced, it is used only as an oracle diagnostic, not as a practical pricing method.

\paragraph{Amortisation}

The task-aligned lift defines a deterministic map $(H,T)\mapsto\{w_j,x_j\}_{j=1}^m$. To avoid any optimisation at pricing time, we solve Eq.~\eqref{eq:tal} on a regular design over $(H,\log T)$ and interpolate the sorted parameters $(\log w_j,\log x_j)$ bilinearly. This gives the amortised task-aligned lift $A_m(H,\log T)$, whose evaluation costs microseconds and involves no per-instance fitting. In the synthetic experiments the design covers the in-distribution domain. For the SPX calibration experiment, calibrated roughness is allowed to vary over the full admissible range explored by the market calibration, so we build a separate market-range amortised lift over $H\in[0.02,0.49]$ and the relevant maturity domain, with $\Pi$ covering the calibration bounds. This prevents the market experiment from relying on extrapolation outside the support of the design.

For a fixed state budget $m$ we compare the tuned geometric quadrature, the BL2 rule (computed with the authors' implementation), BL2$_+$, $L^2_+$, the task-aligned lift, and its amortisation, with the per-instance oracle as a reference point.

\subsection{Why learning can improve pricing}
\label{sec:theoretical_motivation}

Existing Markovian approximations are often justified by 
weak-error estimates. For suitable classes of European 
payoffs, these estimates yield bounds of the form
\begin{equation}
    \left|P_k^{\mathrm{ref}}-P_k^{(m)}\right|
    \leq C_{\mu,k}
    \left\|K_H-K^{(m)}\right\|_{L^1(0,T)}
    ,
    \label{eq:weak_error_bound}
\end{equation}
where $C_{\mu,k}$ depends on the model parameters and 
payoff class \citep{bayer2023weakmarkovianapproximationsrough}.
Such bounds justify controlling global kernel error but do 
not establish that a kernel-optimal approximation is optimal 
for a particular pricing task. In particular, perturbations 
of comparable kernel norm may induce substantially different 
pricing errors. We formalise this distinction by 
characterising the local geometry of the task-aligned 
objective introduced in Section~\ref{sec:tal}

Fix $(H,T)$ and a finite pricing task $\mathcal{T}$ 
consisting of parameter--strike pairs $(\eta,k)$. Let 
$P_{\eta,k}[K]$ denote the COS price obtained from the 
Riccati--Volterra equation with kernel $K$, keeping the 
Fourier frequencies and inversion coefficients fixed for 
each configuration. Write
\(
    \delta K = K^{(m)}-K_H.
\)
For each Fourier frequency $u$, let $h_u$ denote the 
reference Riccati solution associated with $K_H$. Define 
the adjoint state $\chi_u$ through the backward Volterra 
equation
\begin{equation}
\begin{aligned}
    \chi_u(s)
    ={}& \kappa\theta
    + F_x(u,h_u(s))
    \left[
        V_0+
        \int_s^T K_H(t-s)\chi_u(t)\,dt
    \right],
\end{aligned}
\label{eq:adjoint_volterra}
\end{equation}
where $F_x$ denotes the derivative of $F$ with respect 
to its second argument. The corresponding kernel 
sensitivity is
\begin{equation}
    g_u(\tau)
    =
    \int_\tau^T
    \chi_u(t)F(u,h_u(t-\tau))\,dt.
    \label{eq:kernel_sensitivity}
\end{equation}
Intuitively, $g_u(\tau)$ measures the first-order effect 
of perturbing the kernel at lag $\tau$ on the log 
characteristic function. Since COS inversion is linear 
in the characteristic function, the vega-normalised 
pricing sensitivity is
\begin{equation}
    \bar g_{\eta,k}(\tau)
    =
    \frac{1}{\mathcal{V}_{\eta,k}^{\mathrm{ref}}}
    \operatorname{Re}
    \left[
        \sum_{n=0}^{N_{\mathrm{COS}}-1}
        c_{\eta,k,n}
        \varphi_H(u_n,T)g_{u_n}(\tau)
    \right],
    \label{eq:pricing_sensitivity}
\end{equation}
where $c_{\eta,k,n}$ are the kernel-independent COS 
coefficients, and the dependence of the Riccati solutions 
and Fourier frequencies on $\eta$ is suppressed for 
notational convenience.

\begin{proposition}[Local geometry of the pricing-aware loss]
\label{prop:task_geometry}
Assume that the reference Riccati solutions are continuous 
and bounded on $[0,T]$ at the Fourier frequencies used by 
the pricer, and that the reference Black--Scholes vegas 
are strictly positive for all $(\eta,k)\in\mathcal{T}$.
For sufficiently small 
$\|\delta K\|_{L^1(0,T)}$, the empirical task-aligned 
objective satisfies
\begin{equation}
    \mathcal{L}_{\mathcal{T}}(K^{(m)})
    =
    \left\langle
        \delta K,\mathcal{C}_{\mathcal{T}}\delta K
    \right\rangle_{L^2(0,T)}
    +
    O\!\left(\|\delta K\|_{L^1(0,T)}^3\right),
    \label{eq:task_loss_expansion}
\end{equation}
where
\begin{equation}
    \mathcal{L}_{\mathcal{T}}(K)
    =
    \frac{1}{|\mathcal{T}|}
    \sum_{(\eta,k)\in\mathcal{T}}
    \left[
        \frac{
            P_{\eta,k}[K]-P_{\eta,k}[K_H]
        }{
            \mathcal{V}_{\eta,k}^{\mathrm{ref}}
        }
    \right]^2
\quad \text{and} \quad
    \mathcal{C}_{\mathcal{T}}
    =
    \frac{1}{|\mathcal{T}|}
    \sum_{(\eta,k)\in\mathcal{T}}
    \bar g_{\eta,k}\otimes\bar g_{\eta,k}.
    \label{eq:task_sensitivity_operator}
\end{equation}
The operator $\mathcal{C}_{\mathcal{T}}$ is positive 
semidefinite and has rank at most $|\mathcal{T}|$.
\end{proposition}

The proof is given in Appendix~\ref{app:proofs}.

Proposition~\ref{prop:task_geometry} establishes that 
the task-aligned objective locally induces the kernel 
seminorm
\begin{equation}
    \|\delta K\|_{\mathcal{T}}^2
    =
    \frac{1}{|\mathcal{T}|}
    \sum_{(\eta,k)\in\mathcal{T}}
    \left\langle
        \bar g_{\eta,k},\delta K
    \right\rangle_{L^2(0,T)}^2.
    \label{eq:task_seminorm}
\end{equation}
Unlike the global $L^2$ norm, which penalises all kernel 
perturbation directions equally in the underlying Hilbert 
space, this seminorm weights directions according to 
their influence on option prices. Let $(\lambda_\ell,e_\ell)$ 
denote the non-zero eigenpairs of 
$\mathcal{C}_{\mathcal{T}}$, with orthonormal eigenfunctions 
$e_\ell$. Then
\begin{equation}
    \|\delta K\|_{\mathcal{T}}^2
    =
    \sum_{\ell=1}^{r}
    \lambda_\ell
    \left|
        \langle e_\ell,\delta K\rangle_{L^2(0,T)}
    \right|^2,
    \label{eq:task_spectral_decomposition}
\end{equation}
where $r=\operatorname{rank}(\mathcal{C}_{\mathcal{T}})$.
When the eigenvalues decay rapidly, the leading-order 
pricing error is primarily governed by a small number of 
sensitivity directions. Perturbations in the null space 
of $\mathcal{C}_{\mathcal{T}}$ affect the prices of the 
task family only through higher-order terms.

This interpretation extends naturally to the Markovian 
representation. The fractional kernel admits the 
Laplace representation
\begin{equation}
    K_H(\tau)
    =
    \int_0^\infty e^{-x\tau}\Lambda_H(dx),
    \qquad
    \Lambda_H(dx)
    =
    c_H x^{-H-1/2}\,dx,
    \label{eq:task_laplace_representation}
\end{equation}
where 
$c_H=[\Gamma(H+1/2)\Gamma(1/2-H)]^{-1}$.
Define the Laplace-transformed pricing sensitivities
\begin{equation}
    \widehat g_{\eta,k}(x)
    =
    \int_0^T
    \bar g_{\eta,k}(\tau)e^{-x\tau}\,d\tau.
    \label{eq:transformed_sensitivity}
\end{equation}
By Fubini's theorem, the first-order vega-normalised 
pricing error becomes
\begin{equation}
\begin{aligned}
    \langle\bar g_{\eta,k},\delta K\rangle_{L^2(0,T)}
    ={}&
    \sum_{j=1}^{m}
    w_j\widehat g_{\eta,k}(x_j)
    -
    \int_0^\infty
    \widehat g_{\eta,k}(x)\Lambda_H(dx).
\end{aligned}
\label{eq:task_quadrature}
\end{equation}

A task-aligned exponential approximation can therefore
be interpreted as a quadrature rule for $\Lambda_H$,
targeting the family of functions induced by pricing
sensitivities. For positive weights, this corresponds
to a positive quadrature rule; allowing signed weights
relaxes the positivity constraint and enlarges the
feasible approximation space.

This provides a structural explanation for why freely 
optimising the exponential modes can improve pricing at 
small state budgets: an effective pricing representation 
need not reproduce every feature of fractional memory 
equally well, but should prioritise those kernel directions 
that are relevant to the financial task. The nonlinear 
objective of Section~\ref{sec:tal} additionally accounts for effects 
beyond the first-order approximation.

The result is local and does not guarantee that a 
low-rank sensitivity operator admits an equally 
low-dimensional positive exponential representation, 
nor that the non-convex optimisation identifies a 
globally optimal lift. Whether task alignment translates 
into improved state efficiency is the empirical question 
addressed in the following sections.

\section{Setup and Data}
\label{sec:data}

\paragraph{Synthetic configurations}
A configuration is a rough-Heston parameter vector $\mu=(H,\kappa,\theta,\nu,\rho,V_0)$ together with a maturity $T$, and its label is the model's own reference smile. Parameters are drawn uniformly from the boxes of Table~\ref{tab:params}, and maturities log-uniformly. The main family has $150$ configurations and the stress family has $50$. Strikes lie in log-moneyness $k\in[-0.20,0.16]$. The amortisation grid has $11\times13$ points over $H\in[0.04,0.26]$ and $T\in[0.09,2.2]$, slightly wider than the main box, so that no evaluation configuration requires extrapolation.

\paragraph{SPX options}
We use the Options Price Reporting Authority (OPRA) end-of-day SPX option panel from April 2013 to November 2025, which contains calls and puts with vendor implied volatilities, the underlying level and the risk-free rate. We take one surface per week, from the last trading day of each ISO week, which limits the dependence between consecutive surfaces. For each expiry:
\begin{itemize}
  \item the forward is obtained from put--call parity;
  \item we keep out-of-the-money quotes with positive volume, $|k|\le0.16$ and maturities in $[0.06,1.5]$ years;
  \item at most $14$ strikes, spread across moneyness, are retained.
\end{itemize}
A week is kept if it has at least four such maturities, and at most six maturities are used per week. This yields $646$ weekly surfaces, $3{,}636$ smiles and about $50{,}000$ quotes, of which $288$ weeks precede 2019. Calibration minimises vega-weighted squared implied-volatility errors within the bounds of Table~\ref{tab:params}. The market amortisation grid has $12\times10$ points over $H\in[0.02,0.499]$ and $T\in[0.06,1.5]$.

\begin{table}[t]
\centering
\caption{Parameter ranges: sampling boxes of the synthetic families and bounds of the SPX calibration.}
\label{tab:params}
\small
\begin{tabular}{lccc}
\toprule
 & \textbf{Main} & \textbf{Stress} & \textbf{SPX bounds}\\
\midrule
$H$      & $[0.05,0.25]$   & $[0.02,0.05]$   & $[0.01,0.499]$\\
$\kappa$ & $[0.3,3]$       & $[0.3,3]$       & $[0.01,10]$\\
$\theta$ & $[0.02,0.08]$   & $[0.02,0.08]$   & $[0.005,1]$\\
$\nu$    & $[0.2,0.5]$     & $[0.2,0.6]$     & $[0.05,2.5]$\\
$\rho$   & $[-0.9,-0.3]$   & $[-0.99,-0.9]$  & $[-0.99,-0.05]$\\
$V_0$    & $[0.02,0.08]$   & $[0.02,0.08]$   & $[0.005,0.5]$\\
$T$ (years) & $[0.1,2]$    & $[0.03,0.1]$    & $[0.06,1.5]$\\
\bottomrule
\end{tabular}
\end{table}

\section{Experiments}
\label{sec:experiments}

\subsection{Protocol}
\label{sec:protocol}

We compare all lifts at the same state budget $m$, using BL2~\citep{bayer2023weakmarkovianapproximationsrough} as the main baseline. We also consider its positive-weight version (BL2$_+$), the exact $L^2_+$ rule, geometric quadrature, and free-node kernel fits based on $L^1$ and $L^2$ errors. The per-instance oracle serves as a benchmark for what can be achieved when all model parameters are known.

The task-aligned lift depends only on $(H,T)$. The remaining parameters are drawn from the family prior $\Pi$, using separate training and validation sets. Positive and signed variants are compared through validation. Neither the test configurations nor the test strikes are used to build or select the lifts. For the synthetic studies, we use the main and stress families described in Sec.~\ref{sec:data}, training on the corresponding parameter boxes.

Reference prices are computed with the fractional-Adams pricer and independently checked against the dense sinc engine, which also provides the training prices. This allows us to separate pricing errors from numerical uncertainty. Differences between methods are considered resolvable only when they exceed twice the reference uncertainty.

Our main metric is the worst implied-volatility error across test strikes, measured in basis points. We report medians, upper quantiles, bootstrap confidence intervals and paired comparisons against BL2, using the Wilcoxon signed-rank test with Holm--Bonferroni correction. Numerical settings, optimisation details, baseline fallbacks and reference certification procedures are provided in the supplementary material.

Motivated by evidence that aggregate errors can conceal financially relevant conditional failures \citep{cifer}, we complement median accuracy with upper quantiles and stress-family evaluations.

\subsection{Reproducibility}

To increase the replicability of our results, we provide the code and the information to retrieve the data we use in the experiments directly from the provider in \url{https://github.com/arthurchagas1/Learning-Markovian-Lifts-for-Rough-Heston}

\section{Results}
\label{sec:results}

\subsection{RQ1 --- A better kernel approximation does not give better prices}

Table~\ref{tab:rq1} sets kernel error against pricing error on the main family. Our exact $L^2_+$ rule is the most accurate kernel approximation for every $m\ge2$: its kernel error is lower than that of BL2 on every configuration, and about half of it at $m=4$. Yet it prices worse than BL2 on almost every configuration, by a factor that grows with $m$.

The task-aligned lift shows the opposite picture. Its kernel error is essentially the same as that of BL2, but its pricing error is an order of magnitude smaller. The ranking by kernel error and the ranking by pricing error therefore disagree, in both directions. The stress family shows the same pattern.

\begin{table}[t]
\centering
\caption{RQ1: kernel error versus pricing error on the main family. Entries are medians over the $150$ configurations: relative $L^2$ kernel error on $[0,T]$, and worst-strike implied-volatility error on the test strikes (bps).}
\label{tab:rq1}
\small
\begin{tabular}{lcccccc}
\toprule
 & \multicolumn{2}{c}{$m=2$} & \multicolumn{2}{c}{$m=4$} & \multicolumn{2}{c}{$m=6$}\\
\cmidrule(lr){2-3}\cmidrule(lr){4-5}\cmidrule(lr){6-7}
 & kernel & price & kernel & price & kernel & price\\
\midrule
Geometric       & 0.449 & 132   & 0.299 & 80.3   & 0.234 & 55.1\\
L1-opt          & 0.337 & 17.6  & 0.302 & 6.00   & 0.248 & 3.51\\
L2-opt          & 0.315 & 2.84  & 0.264 & 3.88   & 0.227 & 2.69\\
$L^2_+$         & 0.252 & 2.89  & 0.113 & 1.13   & 0.058 & 0.460\\
BL2             & 0.355 & 1.40  & 0.229 & 0.030  & 0.160 & 0.0010\\
Task-aligned    & 0.308 & 0.043 & 0.229 & 0.0026 & 0.160 & 0.0002\\
\bottomrule
\end{tabular}
\end{table}

The rank diagnostic in Table~\ref{tab:rank} explains why. At every $(H,T)$ we examined, the pricing sensitivities have numerical rank $2$, $3$, $4$ and $5$ at tolerances from $10^{-2}$ to $10^{-5}$, and these ranks change with neither $H$ nor $T$. The Laplace operator, which is the family a kernel-norm rule implicitly guards against, needs several times more directions at the same tolerances. A kernel-norm rule therefore spends its few states on directions that the pricing task never sees.

The comparison with random smooth functions shows where the low rank comes from. Functions that vary on the scale of the maturity have ranks close to those of the task, while rougher functions need more directions. The pricing sensitivities are smooth functions of the lag, which is what makes a small number of well-placed states sufficient. By the Tchakaloff argument of Appendix~\ref{app:mathematical_derivations}, a rank of $3$ at tolerance $10^{-3}$ means that four positive states can be exact to first order up to that tolerance.

\begin{table}[t]
\centering
\caption{Numerical rank of the Laplace-transformed sensitivities at relative tolerance $\varepsilon$, under the same measure weighting. Ranges are taken over the $3\times3$ grid $H\in\{0.05,0.15,0.25\}$, $T\in\{0.1,0.5,2\}$. Smooth random functions are Mat\'ern-$5/2$ Gaussian-process draws with correlation length $\ell$, equal in number to the tasks.}
\label{tab:rank}
\small
\begin{tabular}{lcccc}
\toprule
 & $\varepsilon=10^{-2}$ & $10^{-3}$ & $10^{-4}$ & $10^{-5}$\\
\midrule
Pricing task                      & 2 & 3 & 4 & 5\\
Laplace operator (kernel norm)    & 7--10 & 14--16 & 20--23 & 26--29\\
Smooth random, $\ell=T$           & 2 & 3--4 & 5 & 6\\
Smooth random, $\ell=0.3T$        & 3 & 5 & 6--7 & 8--9\\
Smooth random, $\ell=0.1T$        & 4 & 6 & 8--9 & 10--11\\
\bottomrule
\end{tabular}
\end{table}

\subsection{RQ2 --- The task-aligned lift is several times more accurate than BL2 at the same number of states}

Table~\ref{tab:rq2} contains the central result.

\paragraph{Main family}
The task-aligned lift beats BL2 at every budget. The median paired ratio is about two with one state and between ten and twenty-six with two to four states, and the lift wins on every configuration from $m=2$ to $m=4$. With six states both methods are below a thousandth of a basis point, which is close to the reference uncertainty. Only about a third of the configurations then show a resolvable difference, and on all of them the task-aligned lift is the more accurate. Every difference is significant after the Holm correction.

The older constructions are far behind. The tuned geometric quadrature is off by tens of basis points even with six states, and the free-node kernel fits stall at a few basis points. BL2 is clearly the strongest kernel-level baseline.

In terms of states, the task-aligned lift needs about two fewer states than BL2 for the same accuracy. The median number of states needed to reach $0.1$ bp falls from four to two, and the number needed to reach $0.01$ bp falls from six to four.

\paragraph{Stress family}
The same pattern holds, with larger errors overall. The task-aligned lift is about four times more accurate than BL2 with one state, twenty-one times with two, and seven to ten times with three or four. It wins on essentially every configuration. With six states the two methods tie, because their difference is smaller than the reference uncertainty on almost every configuration. To reach $0.1$ bp, BL2 needs six states and the task-aligned lift needs four.

\paragraph{The cost of conditioning only on $(H,T)$}
On the main family, the per-instance oracle is about twice as accurate as the task-aligned lift for $m\le4$, and equally accurate at $m=6$. Most of the attainable gain is therefore obtained without looking at the configuration being priced. On the stress family the oracle is not better than the task-aligned lift. Fitting a single configuration on a handful of strikes appears to be a less stable problem than fitting a family of draws.

\paragraph{The cost of positive weights}
Positivity matters only at larger budgets. Up to three states the signed variant does not help, and the positive variant is selected in most cases. From four states on, validation increasingly prefers the signed variant. At six states the positive lift is no longer competitive: on the main family it is less accurate than BL2, which itself uses negative weights there, while the signed lift is about five times more accurate than BL2. Kernel-level rules pay the same price: forcing the weights of BL2 to be non-negative (BL2$_+$) makes it more than two hundred times less accurate at $m=6$.

\begin{table}[t]
\centering
\caption{RQ2: Median worst-strike IV error on test strikes (bps). Task-aligned: validation-selected $(H,T)$-conditioned lift; Oracle: per-instance fit (reference only). Ratio: median paired BL2/task-aligned error ratio; Win: share favoring task-aligned. Task-aligned outperforms BL2 (Holm-adjusted $p<10^{-4}$) except at stress $m=6$, where the difference is below reference uncertainty (median: $0.0005$ bps main, $0.017$ bps stress).}
\label{tab:rq2}
\small
\setlength{\tabcolsep}{4pt}
\begin{tabular}{rrrrrrrrrrr}
\toprule
$m$ & Geom. & L1-opt & L2-opt & $L^2_+$ & BL2$_+$ & BL2 & Task-al. & Oracle & Ratio & Win\\
\midrule
\multicolumn{11}{l}{\emph{Main family ($150$ configurations)}}\\
1 & 123  & 35.3 & 4.56 & 5.59  & 5.59  & 5.59   & 2.66   & 2.09   & 1.8  & 83\%\\
2 & 132  & 17.6 & 2.84 & 2.89  & 1.40  & 1.40   & 0.043  & 0.021  & 26   & 100\%\\
3 & 88.8 & 4.48 & 3.07 & 1.69  & 0.264 & 0.264  & 0.025  & 0.013  & 10.6 & 100\%\\
4 & 80.3 & 6.00 & 3.88 & 1.13  & 0.042 & 0.030  & 0.0026 & 0.0013 & 11.5 & 100\%\\
6 & 55.1 & 3.51 & 2.69 & 0.460 & 0.278 & 0.0010 & 0.0002 & 0.0002 & 4.6  & 93\%\\
\midrule
\multicolumn{11}{l}{\emph{Stress family ($50$ configurations)}}\\
1 & 469 & 197  & 53.4 & 45.4 & 45.4  & 45.4   & 9.83   & 9.94   & 4.3  & 98\%\\
2 & 447 & 196  & 52.5 & 23.0 & 4.59  & 4.59   & 0.188  & 0.699  & 21.0 & 100\%\\
3 & 413 & 54.5 & 31.2 & 17.0 & 1.02  & 1.02   & 0.110  & 0.212  & 10.3 & 100\%\\
4 & 250 & 44.5 & 30.8 & 14.8 & 0.172 & 0.166  & 0.024  & 0.032  & 7.3  & 100\%\\
6 & 167 & 53.6 & 32.6 & 11.6 & 1.28  & 0.0082 & 0.0087 & 0.051  & 1.0  & 60\%\\
\midrule
\multicolumn{11}{l}{\emph{Median number of states needed} (BL2 $\to$ task-aligned): main $4\to2$ for $0.1$ bp and $6\to4$ for $0.01$ bp;}\\
\multicolumn{11}{l}{\quad stress $4\to2$ for $1$ bp and $6\to4$ for $0.1$ bp.}\\
\bottomrule
\end{tabular}
\end{table}

\subsection{RQ3 --- The amortised lift keeps most of the gain}

Table~\ref{tab:rq3} compares the amortised lift with BL2 and with the lift fitted at each $(H,T)$.

\paragraph{Accuracy of the amortised lift}
The amortised lift beats BL2 at every budget, with median error ratios of approximately $2$, $30$, $6.5$ and $6$ for one to four states, winning on at least $96\%$ of configurations from $m=2$ onwards. It matches the lift fitted at each $(H,T)$ with one and two states, and gives up about a factor of two at three and four while retaining most of the advantage. At four states, its $90$th percentile and worst-case errors are approximately one fifth and one fifteenth of BL2's, respectively.

\begin{table}[t]
\centering
\caption{RQ3: amortised lift on the main family (worst-strike implied-volatility error, bps). Medians are given, plus the $90$th percentile (q90) for BL2 and for the amortised lift. Naive interpolates log-weights and log-rates directly. Ratio and Win compare the amortised lift with BL2. Every comparison has Holm-adjusted $p<10^{-4}$.}
\label{tab:rq3}
\small
\begin{tabular}{rcccccccc}
\toprule
 & \multicolumn{4}{c}{median} & & & \multicolumn{2}{c}{q90}\\
\cmidrule(lr){2-5}\cmidrule(lr){8-9}
$m$ & BL2 & fitted at $(H,T)$ & naive table & amortised & Ratio & Win & BL2 & amortised\\
\midrule
1 & 5.59  & 2.66   & 2.64  & 2.67   & 1.8 & 84\%  & 10.9  & 5.68\\
2 & 1.40  & 0.043  & 0.048 & 0.038  & 30  & 100\% & 3.93  & 0.115\\
3 & 0.264 & 0.025  & 0.255 & 0.039  & 6.5 & 99\%  & 0.484 & 0.108\\
4 & 0.030 & 0.0026 & 0.830 & 0.0053 & 5.9 & 96\%  & 0.077 & 0.016\\
\bottomrule
\end{tabular}
\end{table}

\subsection{RQ4 --- On SPX, the task-aligned lift is the more reliable engine from two states on}

\paragraph{The data prefer mild roughness}
Wider calibration bounds reduce the share of weeks with parameters at a bound from four in five to one in three, with most remaining cases approaching $H=\tfrac12$. The calibrated roughness has a median of $0.42$ and ranges from $0.17$ to nearly $\tfrac12$. Fixing $H$ near $0.1$ increases the median root mean squared error (RMSE) by about ten basis points, and only one week in eight can be fitted within one basis point of its optimum with $H\le0.12$. The market study therefore covers both rough and nearly Markovian regimes.

\paragraph{State efficiency on real smiles}
The certified reference has a median uncertainty of $0.006$ bps per week, independently cross-checked by the sinc engine; fewer than $0.5\%$ of strikes are excluded. Table~\ref{tab:rq4} shows that neither lift is adequate with one state. From two states on, the task-aligned lift improves upon BL2 across medians, upper quantiles and accuracy rates on the full sample. With two states it is about ten times more accurate, while with four it prices $98\%$ of weeks within one basis point, compared with $49\%$ for BL2. Geometric quadrature produces errors of hundreds of basis points and is omitted.

\begin{table}[t]
\centering
\caption{RQ4: Worst-strike implied-volatility error (bps) across weekly SPX surfaces, maximised over maturities. Ratio denotes the median paired BL2/task-aligned error ratio, Win the task-aligned win rate, and $\leq 1$ bp the accuracy rate (BL2/task-aligned). Rough weeks have $H<0.3$; BL2 succeeded excludes fallback rules. Holm-adjusted $p<10^{-4}$ for all $m\geq2$, except the last row ($p=0.12$). Median reference uncertainty: $0.006$ bps.}
\label{tab:rq4}
\small
\setlength{\tabcolsep}{4pt}
\begin{tabular}{rrrrrrrrr}
\toprule
 & \multicolumn{3}{c}{median} & & & \multicolumn{2}{c}{q90} & \\
\cmidrule(lr){2-4}\cmidrule(lr){7-8}
$m$ & BL2$_+$ & BL2 & Task-al. & Ratio & Win & BL2 & Task-al. & $\le1$ bp\\
\midrule
\multicolumn{9}{l}{\emph{All weeks ($646$)}}\\
1 & 3.39  & 3.39  & 4.38  & 1.0  & 52\%  & 12.3 & 9.33  & 34 / 19\%\\
2 & 2.48  & 4.54  & 0.498 & 9.6  & 84\%  & 17.7 & 1.54  & 15 / 74\%\\
3 & 0.564 & 0.806 & 0.273 & 4.5  & 80\%  & 80.6 & 0.905 & 52 / 91\%\\
4 & 2.18  & 1.12  & 0.123 & 3.9  & 75\%  & 42.6 & 0.553 & 49 / 98\%\\
\midrule
\multicolumn{9}{l}{\emph{Rough weeks ($135$)}}\\
1 & 11.9  & 11.9  & 7.52  & 1.5  & 90\%  & 22.8 & 14.4  & 0 / 0\%\\
2 & 7.24  & 7.24  & 0.445 & 14.7 & 100\% & 10.0 & 1.47  & 0 / 77\%\\
3 & 0.528 & 0.528 & 0.216 & 2.3  & 87\%  & 1.05 & 0.623 & 87 / 93\%\\
4 & 2.37  & 0.060 & 0.042 & 1.4  & 72\%  & 0.221 & 0.188 & 97 / 97\%\\
\bottomrule
\end{tabular}
\end{table}

\paragraph{Where BL2 breaks down}
As $H$ approaches $\tfrac12$, BL2 frequently fails to return a rule or produces large, alternating-sign weights, resulting in substantial pricing errors. However, numerical instability does not fully explain the gains from task alignment. On rough weeks, the task-aligned lift outperforms BL2 at every budget, by nearly fifteen times with two states. Restricting the comparison to weeks where BL2 succeeds, the advantage persists with two and three states. With four states, BL2 achieves a lower median error, although the difference is not significant and its upper quantiles remain substantially worse.

\paragraph{The gains persist out of time and in calibration}
Trained exclusively on pre-2019 parameters, the amortised lift maintains its accuracy and advantage over BL2 on later weeks and during periods of elevated implied volatility. Recalibrating with four states recovers the reference parameters to about a thousandth and, when evaluated with the reference pricer, achieves a market fit within one basis point of the reference calibration on $94\%$ of weeks. Median calibration is four times faster, despite the reference engine already using a coarse grid during optimisation.

\section{Conclusions and Limitations}
\label{sec:conclusion}

We showed that the efficiency of Markovian approximations for rough-Heston pricing depends not only on how accurately they reproduce fractional memory, but also on how they represent the kernel directions relevant to pricing. Our local sensitivity analysis explains this distinction through a quadrature interpretation, while the experiments demonstrate its practical relevance. Task-aligned lifts achieve substantially lower implied-volatility errors than BL2 at matched state budgets, typically requiring two fewer states to reach comparable accuracy. The amortised construction preserves most of these gains without online optimisation and accelerates weekly SPX calibration fourfold.

These findings remain limited to European options under rough Heston. The theoretical guarantees are local, and low numerical rank alone does not guarantee the existence or successful optimisation of an equally compact positive lift. Extending task alignment to other Volterra models, path-dependent derivatives and stronger theoretical guarantees constitutes a natural direction for future research. More broadly, our results suggest that learning task-specific numerical representations within established financial models can be more effective than optimising their global approximation properties.  While signed approximations can improve numerical pricing accuracy, they do not automatically preserve the stochastic admissibility of positive Markovian lifts. Moreover, maturity-dependent approximations need not define a single stochastic process across all maturities. Our conclusions therefore concern European option pricing and calibration.

\bibliography{rQUFguide}

@String{Computing = "Computing" }

@String{Springer = "Springer-Verlag" }

@ArtifactSoftware{R,
    title = {R: A Language and Environment for Statistical Computing},
    author = {{R Core Team}},
    organization = {R Foundation for Statistical Computing},
    address = {Vienna, Austria},
    year = {2019},
    url = {https://www.R-project.org/},
}

@article{gatheral2018volatilityrough,
author = {Jim Gatheral and Thibault Jaisson and Mathieu Rosenbaum},
title = {Volatility is rough},
journal = {Quantitative Finance},
volume = {18},
number = {6},
pages = {933--949},
year = {2018},
publisher = {Routledge},
doi = {10.1080/14697688.2017.1393551},


URL = { 
    
        https://doi.org/10.1080/14697688.2017.1393551
    
    

},
eprint = { 
    
        https://doi.org/10.1080/14697688.2017.1393551
    
    

}

}

@article{bayer2016pricingrough,
author = {Christian Bayer and Peter Friz and Jim Gatheral},
title = {Pricing under rough volatility},
journal = {Quantitative Finance},
volume = {16},
number = {6},
pages = {887--904},
year = {2016},
publisher = {Routledge},
doi = {10.1080/14697688.2015.1099717},


URL = { 
    
        https://doi.org/10.1080/14697688.2015.1099717
    
    

},
eprint = { 
    
        https://doi.org/10.1080/14697688.2015.1099717
    
    

}

}

@misc{bayer2019deepcalibration,
      title={On deep calibration of (rough) stochastic volatility models}, 
      author={Christian Bayer and Blanka Horvath and Aitor Muguruza and Benjamin Stemper and Mehdi Tomas},
      year={2019},
      eprint={1908.08806},
      archivePrefix={arXiv},
      primaryClass={q-fin.MF},
      url={https://arxiv.org/abs/1908.08806}, 
}

@article{heston1993,
 ISSN = {08939454, 14657368},
 URL = {http://www.jstor.org/stable/2962057},
 author = {Steven L. Heston},
 journal = {The Review of Financial Studies},
 number = {2},
 pages = {327--343},
 publisher = {[Oxford University Press, Society for Financial Studies]},
 title = {A Closed-Form Solution for Options with Stochastic Volatility with Applications to Bond and Currency Options},
 urldate = {2026-06-17},
 volume = {6},
 year = {1993}
}

@article{eleuch,
author = {El Euch, Omar and Rosenbaum, Mathieu},
title = {The characteristic function of rough Heston models},
journal = {Mathematical Finance},
volume = {29},
number = {1},
pages = {3-38},
doi = {https://doi.org/10.1111/mafi.12173},
url = {https://onlinelibrary.wiley.com/doi/abs/10.1111/mafi.12173},
eprint = {https://onlinelibrary.wiley.com/doi/pdf/10.1111/mafi.12173},
year = {2019}
}

@article{abijaber,
author = {Abi Jaber, Eduardo and El Euch, Omar},
title = {Multifactor Approximation of Rough Volatility Models},
journal = {SIAM Journal on Financial Mathematics},
volume = {10},
number = {2},
pages = {309-349},
year = {2019},
doi = {10.1137/18M1170236},

URL = { 
    
        https://doi.org/10.1137/18M1170236
    
    

},
eprint = { 
    
        https://doi.org/10.1137/18M1170236
    
    

}
}

@misc{bayer2023weakmarkovianapproximationsrough,
      title={Weak Markovian Approximations of Rough Heston}, 
      author={Christian Bayer and Simon Breneis},
      year={2023},
      eprint={2309.07023},
      archivePrefix={arXiv},
      primaryClass={q-fin.CP},
      url={https://arxiv.org/abs/2309.07023}, 
}

@Article{bennedsen2017hybrid,
author={Bennedsen, Mikkel
and Lunde, Asger
and Pakkanen, Mikko S.},
title={Hybrid scheme for Brownian semistationary processes},
journal={Finance and Stochastics},
year={2017},
month={Oct},
day={01},
volume={21},
number={4},
pages={931-965},
issn={1432-1122},
doi={10.1007/s00780-017-0335-5},
url={https://doi.org/10.1007/s00780-017-0335-5}
}

@article{mccrickerd2018turbocharging,
author = {Ryan McCrickerd and Mikko S. Pakkanen},
title = {Turbocharging Monte Carlo pricing for the rough Bergomi model},
journal = {Quantitative Finance},
volume = {18},
number = {11},
pages = {1877--1886},
year = {2018},
publisher = {Routledge},
doi = {10.1080/14697688.2018.1459812},


URL = { 
    
        https://doi.org/10.1080/14697688.2018.1459812
    
    

},
eprint = { 
    
        https://doi.org/10.1080/14697688.2018.1459812
    
    

}

}

@Article{cuchiero2019markovian,
author={Cuchiero, Christa
and Teichmann, Josef},
title={Markovian lifts of positive semidefinite affine Volterra-type processes},
journal={Decisions in Economics and Finance},
year={2019},
month={Dec},
day={01},
volume={42},
number={2},
pages={407-448},
issn={1129-6569},
doi={10.1007/s10203-019-00268-5},
url={https://doi.org/10.1007/s10203-019-00268-5}
}

@ARTICLE{horvath2021deeplearningvolatility,
title = {Deep learning volatility: a deep neural network perspective on pricing and calibration in (rough) volatility models},
author = {Horvath, Blanka and Muguruza, Aitor and Tomas, Mehdi},
year = {2021},
journal = {Quantitative Finance},
volume = {21},
number = {1},
pages = {11-27},
url = {https://EconPapers.repec.org/RePEc:taf:quantf:v:21:y:2021:i:1:p:11-27}
}

@article{ferguson2018deeply, title={Deeply learning derivatives}, author={Ferguson, Ryan and Green, Andrew}, journal={arXiv preprint arXiv:1809.02233}, year={2018} }

@article{sirignano2018dgm,
title = {DGM: A deep learning algorithm for solving partial differential equations},
journal = {Journal of Computational Physics},
volume = {375},
pages = {1339-1364},
year = {2018},
issn = {0021-9991},
doi = {https://doi.org/10.1016/j.jcp.2018.08.029},
url = {https://www.sciencedirect.com/science/article/pii/S0021999118305527},
author = {Justin Sirignano and Konstantinos Spiliopoulos}
}

@misc{raissi2019physics,
      title={Physics Informed Deep Learning (Part I): Data-driven Solutions of Nonlinear Partial Differential Equations}, 
      author={Maziar Raissi and Paris Perdikaris and George Em Karniadakis},
      year={2017},
      eprint={1711.10561},
      archivePrefix={arXiv},
      primaryClass={cs.AI},
      url={https://arxiv.org/abs/1711.10561}, 
}

@article{lu2019deeponet,
   title={Learning nonlinear operators via DeepONet based on the universal approximation theorem of operators},
   volume={3},
   ISSN={2522-5839},
   url={http://dx.doi.org/10.1038/s42256-021-00302-5},
   DOI={10.1038/s42256-021-00302-5},
   number={3},
   journal={Nature Machine Intelligence},
   publisher={Springer Science and Business Media LLC},
   author={Lu, Lu and Jin, Pengzhan and Pang, Guofei and Zhang, Zhongqiang and Karniadakis, George Em},
   year={2021},
   month=Mar, pages={218–229} }

@inproceedings{
li2021fourier,
title={Fourier Neural Operator for Parametric Partial Differential Equations},
author={Zongyi Li and Nikola Borislavov Kovachki and Kamyar Azizzadenesheli and Burigede liu and Kaushik Bhattacharya and Andrew Stuart and Anima Anandkumar},
booktitle={International Conference on Learning Representations},
year={2021},
url={https://openreview.net/forum?id=c8P9NQVtmnO}
}

@article{bayer,
author = {Bayer, Christian and Breneis, Simon},
title = {Markovian approximations of stochastic Volterra equations with the fractional kernel},
journal = {Quantitative Finance},
volume = {23},
number = {1},
pages = {53--70},
year = {2023},
publisher = {Routledge},
doi = {10.1080/14697688.2022.2139193},


URL = { 
    
        https://doi.org/10.1080/14697688.2022.2139193
    
    

},
eprint = { 
    
        https://doi.org/10.1080/14697688.2022.2139193
    
    

}

}

@article{bayer2006tchakaloff,
 ISSN = {00029939, 10886826},
 URL = {http://www.jstor.org/stable/4098162},
 author = {Christian Bayer and Josef Teichmann},
 journal = {Proceedings of the American Mathematical Society},
 number = {10},
 pages = {3035--3040},
 publisher = {American Mathematical Society},
 title = {The Proof of Tchakaloff's Theorem},
 urldate = {2026-11-26},
 volume = {134},
 year = {2006}
}

@INPROCEEDINGS{cifer,
  author={Chagas, Arthur and Bento, Pedro and Aquino, Yan and Buzelin, Arthur and Meira, Wagner and Valle, Cristiano Arbex},
  booktitle={2026 IEEE Symposium on Computational Intelligence for Financial Engineering and Economics (CIFEr)}, 
  title={Latent-Regime Bias Auditing for Volatility Forecasting}, 
  year={2026},
  volume={},
  number={},
  pages={236-243},
  doi={10.1109/CIFEr67845.2026.11692382}}
\bibliographystyle{apalike}

\newpage

\appendix

\section{Proofs}
\label{app:proofs}

\subsection{Proof of Proposition~\ref{prop:task_geometry}}
\label{app:proof_task_geometry}

Fix $(\eta,k)\in\mathcal{T}$ and a Fourier frequency $u$. 
For notational convenience, write
\[
    h=h_u,\qquad
    f(t)=F(u,h(t)),\qquad
    \beta(t)=F_x(u,h(t)).
\]
Let $\widetilde h$ denote the Riccati solution associated 
with $K_H+\delta K$, and set
\[
    e=\widetilde h-h,
    \qquad
    \varepsilon=\|\delta K\|_{L^1(0,T)}.
\]

\paragraph{Local stability of the Riccati solution.}

By assumption, the reference solution $h$ is continuous 
and bounded on $[0,T]$. Choose $R>\|h\|_\infty$ and let 
$\pi_R$ denote the radial projection onto the closed 
complex disc of radius $R$. The truncated nonlinearity
\(
    F_R(u,z)=F(u,\pi_R(z))
\)
is globally Lipschitz in $z$, with Lipschitz constant 
$L_R<\infty$.

For $\gamma>0$, introduce the exponentially weighted 
supremum norm
\(
    \|z\|_\gamma
    =
    \sup_{0\leq t\leq T}
    e^{-\gamma t}|z(t)|.
\)
Since $K_H\in L^1(0,T)$, we may choose $\gamma$ 
sufficiently large that
\[
    L_R\int_0^T e^{-\gamma s}K_H(s)\,ds
    \leq \frac14.
\]
For sufficiently small $\varepsilon$,
\begin{equation}
\begin{aligned}
    &L_R\int_0^T e^{-\gamma s}
    |K_H(s)+\delta K(s)|\,ds
    \space\leq
    L_R\int_0^T e^{-\gamma s}K_H(s)\,ds
    +L_R\varepsilon
    \leq \frac12.
\end{aligned}
\end{equation}
Consequently, the truncated Riccati--Volterra map
\(
    z\longmapsto (K_H+\delta K)*F_R(u,z)
\)
is a contraction in the weighted supremum norm and 
admits a unique continuous fixed point $\widetilde h_R$.

Subtracting the reference equation from the truncated 
perturbed equation gives
\begin{equation}
\begin{aligned}
    \widetilde h_R-h
    ={}&
    \delta K*f
    +
    (K_H+\delta K)*
    \left[
        F_R(u,\widetilde h_R)-F_R(u,h)
    \right].
\end{aligned}
\end{equation}
The contraction estimate and Young's inequality yield
\(
    \|\widetilde h_R-h\|_\gamma
    \leq
    \varepsilon\|f\|_\infty
    +\frac12\|\widetilde h_R-h\|_\gamma.
\)
Hence
\(
    \|\widetilde h_R-h\|_\infty
    \leq
    2e^{\gamma T}\|f\|_\infty\varepsilon.
\)
For sufficiently small $\varepsilon$, the perturbed 
solution remains inside the disc of radius $R$, so the 
truncation is inactive. It therefore solves the original 
Riccati--Volterra equation. In particular,
\begin{equation}
    \|e\|_\infty=O(\varepsilon).
    \label{eq:proof_riccati_stability}
\end{equation}

\paragraph{First-order expansion}

Since $F$ is quadratic in its second argument,
\begin{equation}
    F(u,h+e)
    =
    f+\beta e+\frac{\nu^2}{2}e^2.
    \label{eq:proof_taylor_riccati}
\end{equation}
Subtracting the reference Riccati equation from the 
perturbed equation and using 
Eq.~\eqref{eq:proof_taylor_riccati}, we obtain
\begin{equation}
    e
    =
    \delta K*f
    +K_H*(\beta e)
    +r,
    \label{eq:proof_linearised_riccati}
\end{equation}
where
\begin{equation}
    r
    =
    \delta K*(\beta e)
    +
    \frac{\nu^2}{2}(K_H+\delta K)*e^2.
\end{equation}
By Eq.~\eqref{eq:proof_riccati_stability} and Young's 
inequality,
\begin{equation}
\begin{aligned}
    \|r\|_\infty
    \leq{}&
    \varepsilon\|\beta\|_\infty\|e\|_\infty
    +
    \frac{\nu^2}{2}
    \left(\|K_H\|_{L^1}+\varepsilon\right)
    \|e\|_\infty^2
    ={} O(\varepsilon^2).
\end{aligned}
\label{eq:proof_riccati_remainder}
\end{equation}

Let $\ell_u[K]$ denote the log characteristic function 
associated with kernel $K$. Expanding its defining 
expression gives
\begin{equation}
\begin{aligned}
    \Delta\ell_u
    :={}&
    \ell_u[K_H+\delta K]-\ell_u[K_H]\\
    ={}&
    \int_0^T
    [\kappa\theta+V_0\beta(t)]e(t)\,dt
    +
    \frac{V_0\nu^2}{2}
    \int_0^T e(t)^2\,dt.
\end{aligned}
\end{equation}
It follows that
\begin{equation}
    \Delta\ell_u
    =
    \int_0^T a(t)e(t)\,dt
    +O(\varepsilon^2),
    \qquad
    a(t)=\kappa\theta+V_0\beta(t).
    \label{eq:proof_log_cf_expansion}
\end{equation}

\paragraph{Adjoint representation}

Consider the backward Volterra equation
\begin{equation}
    \chi(s)
    =
    a(s)
    +
    \beta(s)\int_s^T K_H(t-s)\chi(t)\,dt.
    \label{eq:proof_adjoint}
\end{equation}
Reversing time transforms this equation into a linear 
Volterra equation with continuous bounded coefficients 
and an integrable kernel. The same exponentially weighted 
norm argument gives existence and uniqueness of a 
continuous bounded solution.

Multiplying Eq.~\eqref{eq:proof_adjoint} by $e(s)$, 
integrating, and applying Fubini's theorem yields
\begin{equation}
\begin{aligned}
    \int_0^T a(s)e(s)\,ds
    ={}&
    \int_0^T \chi(t)e(t)\,dt
    -
    \int_0^T
    \chi(t)[K_H*(\beta e)](t)\,dt.
\end{aligned}
\end{equation}
Substituting Eq.~\eqref{eq:proof_linearised_riccati} 
and using Eq.~\eqref{eq:proof_riccati_remainder}, 
we obtain
\begin{equation}
\begin{aligned}
    \int_0^T a(t)e(t)\,dt
    ={}&
    \int_0^T
    \chi(t)(\delta K*f)(t)\,dt
    +O(\varepsilon^2).
\end{aligned}
\end{equation}
Exchanging the order of integration once more,
\begin{equation}
\begin{aligned}
    \int_0^T \chi(t)(\delta K*f)(t)\,dt
    \space=
    \int_0^T
    \delta K(\tau)
    \left[
        \int_\tau^T\chi(t)f(t-\tau)\,dt
    \right]d\tau
    \space=
    \int_0^T g_u(\tau)\delta K(\tau)\,d\tau,
\end{aligned}
\end{equation}
where $g_u$ is precisely the kernel sensitivity defined 
in Eq.~\eqref{eq:kernel_sensitivity}. All exchanges 
of integration are justified by boundedness of $\chi$ 
and $f$, and integrability of $\delta K$.

Combining this identity with 
Eq.~\eqref{eq:proof_log_cf_expansion} gives
\begin{equation}
    \Delta\ell_u
    =
    \int_0^T g_u(\tau)\delta K(\tau)\,d\tau
    +O(\varepsilon^2).
    \label{eq:proof_log_cf_sensitivity}
\end{equation}
In particular, boundedness of $g_u$ implies 
$\Delta\ell_u=O(\varepsilon)$.

\paragraph{Propagation through the pricing functional}

Since
\(
    \varphi[K_H+\delta K](u,T)
    =
    \varphi_H(u,T)\exp(\Delta\ell_u),
\)
the expansion of the exponential and
Eq.~\eqref{eq:proof_log_cf_sensitivity} imply
\begin{equation}
\begin{aligned}
    \varphi[K_H+\delta K](u,T)-\varphi_H(u,T)
    \space=
    \varphi_H(u,T)
    \int_0^T g_u(\tau)\delta K(\tau)\,d\tau
    +O(\varepsilon^2).
\end{aligned}
\end{equation}

The finite COS pricing functional is
\[
    P_{\eta,k}[K]
    =
    \operatorname{Re}
    \left[
        \sum_{n=0}^{N_{\mathrm{COS}}-1}
        c_{\eta,k,n}\varphi[K](u_n,T)
    \right].
\]
Because the Fourier frequencies and coefficients are 
fixed, substituting the characteristic-function expansion 
and dividing by the strictly positive reference vega 
gives
\begin{equation}
\begin{aligned}
    &\frac{
        P_{\eta,k}[K_H+\delta K]
        -P_{\eta,k}[K_H]
    }{
        \mathcal{V}_{\eta,k}^{\mathrm{ref}}
    }
    \space=
    \int_0^T
    \bar g_{\eta,k}(\tau)\delta K(\tau)\,d\tau
    +O(\varepsilon^2),
\end{aligned}
\label{eq:proof_normalised_price_expansion}
\end{equation}
where $\bar g_{\eta,k}$ is defined in 
Eq.~\eqref{eq:pricing_sensitivity}. Since the task 
family and the set of Fourier frequencies are finite, 
the remainder constants can be chosen uniformly over 
the task family.

\paragraph{Local geometry of the task objective}

For $H>0$, the fractional kernel belongs to 
$L^2(0,T)$, as does every finite positive SOE. Moreover, 
the pricing sensitivities are bounded and therefore 
belong to $L^2(0,T)$. We may thus express 
Eq.~\eqref{eq:proof_normalised_price_expansion} as
\begin{equation}
    \frac{
        P_{\eta,k}[K^{(m)}]-P_{\eta,k}[K_H]
    }{
        \mathcal{V}_{\eta,k}^{\mathrm{ref}}
    }
    =
    \langle\bar g_{\eta,k},\delta K\rangle_{L^2}
    +O(\varepsilon^2).
\end{equation}
Furthermore,
\(
    \left|
        \langle\bar g_{\eta,k},\delta K\rangle_{L^2}
    \right|
    \leq
    \|\bar g_{\eta,k}\|_\infty\varepsilon
    =
    O(\varepsilon).
\)
Squaring the normalised residual therefore yields
\begin{equation}
\begin{aligned}
    \left[
        \frac{
            P_{\eta,k}[K^{(m)}]-P_{\eta,k}[K_H]
        }{
            \mathcal{V}_{\eta,k}^{\mathrm{ref}}
        }
    \right]^2
    \space=
    \langle\bar g_{\eta,k},\delta K\rangle_{L^2}^2
    +O(\varepsilon^3).
\end{aligned}
\end{equation}
Averaging over $\mathcal{T}$ gives
\begin{equation}
\begin{aligned}
    \mathcal{L}_{\mathcal{T}}(K^{(m)})
    ={}
    \frac{1}{|\mathcal{T}|}
    \sum_{(\eta,k)\in\mathcal{T}}
    \langle\bar g_{\eta,k},\delta K\rangle_{L^2}^2
    +O(\varepsilon^3)
    ={}
    \langle
        \delta K,\mathcal{C}_{\mathcal{T}}\delta K
    \rangle_{L^2}
    +O(\varepsilon^3),
\end{aligned}
\end{equation}
where
\[
    \mathcal{C}_{\mathcal{T}}
    =
    \frac{1}{|\mathcal{T}|}
    \sum_{(\eta,k)\in\mathcal{T}}
    \bar g_{\eta,k}\otimes\bar g_{\eta,k}.
\]
Here, for $g,v\in L^2(0,T)$, the rank-one operator 
$g\otimes g$ acts as
\(
    (g\otimes g)v=g\langle g,v\rangle_{L^2}.
\)
The operator $\mathcal{C}_{\mathcal{T}}$ is self-adjoint 
and positive semidefinite, since for every real-valued 
$v\in L^2(0,T)$,
\[
    \langle v,\mathcal{C}_{\mathcal{T}}v\rangle_{L^2}
    =
    \frac{1}{|\mathcal{T}|}
    \sum_{(\eta,k)\in\mathcal{T}}
    \langle\bar g_{\eta,k},v\rangle_{L^2}^2
    \geq 0.
\]
Finally,
\(
    \operatorname{Im}(\mathcal{C}_{\mathcal{T}})
    \subseteq
    \operatorname{span}
    \left\{
        \bar g_{\eta,k}:(\eta,k)\in\mathcal{T}
    \right\},
\)
which implies
\(
    \operatorname{rank}(\mathcal{C}_{\mathcal{T}})
    \leq|\mathcal{T}|.
\)
This completes the proof.
\hfill$\square$

\section{Complete Mathematical Derivations}
\label{app:mathematical_derivations}

This section complements the proof of Proposition~\ref{prop:task_geometry} by connecting the proposed task-aligned lift to existing results on Markovian approximation, weak pricing errors and positive quadrature rules.

\subsection{From kernel error to pricing error}

The Markovian approximations of \citet{abijaber} and the weak-error analysis of \citet{bayer2023weakmarkovianapproximationsrough} justify approximating rough memory through finite sums of exponentials. Under suitable assumptions on the kernels and payoffs, pricing errors can be controlled by the $L^1$ kernel error. For payoff classes admitting a linear bound, this takes the form
\begin{equation}
    |P_{\eta,k}[K^{(m)}]-P_{\eta,k}[K_H]|
    \leq C_{\eta,k}\|K^{(m)}-K_H\|_{L^1}.
\end{equation}

These bounds provide sufficient conditions for convergence, but do not imply that the lift with the smallest kernel error also gives the most accurate prices. To see why, let $\delta K=K^{(m)}-K_H$ and $\varepsilon=\|\delta K\|_{L^1}$. From the expansion proved in Appendix~\ref{app:proofs}, we have
\begin{equation}
    \frac{P_{\eta,k}[K^{(m)}]-P_{\eta,k}[K_H]}
    {\mathcal{V}_{\eta,k}^{\mathrm{ref}}}
    =
    \langle\bar g_{\eta,k},\delta K\rangle_{L^2}
    +O(\varepsilon^2).
\end{equation}

By H\"older's inequality,
\begin{equation}
    |\langle\bar g_{\eta,k},\delta K\rangle_{L^2}|
    \leq
    \|\bar g_{\eta,k}\|_\infty\|\delta K\|_{L^1}.
\end{equation}

This recovers a local linear kernel-error bound for the fixed COS pricing functional. However, the bound discards information about the direction of $\delta K$. Two perturbations with similar kernel norms may have very different inner products with the pricing sensitivity $\bar g_{\eta,k}$. Our objective retains this information by directly minimising the squared pricing residuals.

Furthermore, since the derivative of the Black--Scholes price with respect to implied volatility is the vega, the inverse function theorem gives
\begin{equation}
    \sigma_{\eta,k}^{(m)}-\sigma_{\eta,k}^{\mathrm{ref}}
    =
    \frac{P_{\eta,k}[K^{(m)}]-P_{\eta,k}[K_H]}
    {\mathcal{V}_{\eta,k}^{\mathrm{ref}}}
    +O(\varepsilon^2),
\end{equation}
provided the reference price lies in the interior of the Black--Scholes price range and its vega is positive. This justifies the vega-normalised objective in Eq.~\eqref{eq:tal}.

\subsection{Task alignment as a quadrature problem}

The fractional kernel and its Markovian approximation can be written as Laplace transforms of continuous and discrete measures, respectively~\citep{abijaber}. Using Eq.~\eqref{eq:task_laplace_representation}, we have
\begin{equation}
    \delta K(\tau)
    =
    \int_0^\infty e^{-x\tau}
    (\Lambda_m-\Lambda_H)(dx),
    \qquad
    \Lambda_m=\sum_{j=1}^m w_j\delta_{x_j}.
\end{equation}

Substituting this representation into the leading-order pricing error and exchanging the order of integration gives
\begin{equation}
\begin{aligned}
    \langle\bar g_{\eta,k},\delta K\rangle_{L^2}
    &=
    \sum_{j=1}^m w_j\widehat g_{\eta,k}(x_j)\\
    &\quad-
    \int_0^\infty\widehat g_{\eta,k}(x)\Lambda_H(dx),
\end{aligned}
\end{equation}
where $\widehat g_{\eta,k}$ is defined in Eq.~\eqref{eq:transformed_sensitivity}. The exchange is justified by the boundedness of the sensitivities and the integrability of $K_H$ on $[0,T]$.

For fixed nodes, define the matrix $B$ and vector $y$ by
\begin{equation}
    B_{ij}=\widehat g_i(x_j),
    \qquad
    y_i=\int_0^\infty\widehat g_i(x)\Lambda_H(dx),
\end{equation}
where $i$ indexes the parameter--strike pairs in $\mathcal{T}$. Proposition~\ref{prop:task_geometry} then gives
\begin{equation}
    \mathcal{L}_{\mathcal{T}}(K^{(m)})
    =
    \frac{1}{|\mathcal{T}|}\|Bw-y\|_2^2
    +O(\varepsilon^3).
\end{equation}

Thus, to first order, learning the lift amounts to finding a quadrature rule that integrates the pricing sensitivities accurately. Unlike conventional kernel approximation, the objective is not to reproduce the fractional kernel equally well in every direction. The nonlinear optimisation in Eq.~\eqref{eq:tal} additionally accounts for higher-order pricing effects.

\subsection{Low-dimensional sensitivities and state efficiency}

The quadrature interpretation connects our method to the generalised Tchakaloff theorem~\citep{bayer2006tchakaloff}, which establishes the existence of positive quadrature rules for finite-dimensional families of integrable functions.

Since $\Lambda_H$ has infinite total mass, we introduce the finite positive measure
\begin{equation}
    \mu_{H,T}(dx)=q_T(x)\Lambda_H(dx),
    \qquad
    q_T(x)=\frac{1-e^{-xT}}{x}.
\end{equation}
Its total mass is
\begin{equation}
    M_{H,T}
    =
    \int_0^\infty\mu_{H,T}(dx)
    =
    \frac{T^{H+1/2}}{\Gamma(H+3/2)}.
\end{equation}

Define $f_i(x)=\widehat g_i(x)/q_T(x)$. These functions are bounded because $|\widehat g_i(x)|\leq\|\bar g_i\|_\infty q_T(x)$. Suppose that they span a space of dimension $r$, with basis $p_1,\ldots,p_r$. Applying the generalised Tchakaloff theorem to $\mu_{H,T}$ and the functions $\{1,p_1,\ldots,p_r\}$ yields positive coefficients $a_j$ and nodes $x_j>0$, with $m\leq r+1$, such that
\begin{equation}
    \sum_{j=1}^m a_jf_i(x_j)
    =
    \int_0^\infty f_i(x)\mu_{H,T}(dx)
\end{equation}
for every task $i$, while preserving the total mass $\sum_j a_j=M_{H,T}$.

Setting $w_j=a_j/q_T(x_j)$ gives
\begin{equation}
    \sum_{j=1}^m w_j\widehat g_i(x_j)
    =
    \int_0^\infty\widehat g_i(x)\Lambda_H(dx).
\end{equation}

Hence, a positive lift with at most $r+1$ states can cancel the first-order pricing error of the entire task family. If this lift also lies within the local stability neighbourhood of Proposition~\ref{prop:task_geometry}, the remaining pricing residual is of order $O(\varepsilon^2)$ and the task loss is of order $O(\varepsilon^4)$.

More generally, suppose that each $f_i$ can be approximated by a function in an $r$-dimensional space with uniform error at most $\delta$. A positive quadrature that integrates this space exactly and preserves the total mass then satisfies
\begin{equation}
    |\langle\bar g_i,\delta K\rangle_{L^2}|
    \leq 2M_{H,T}\delta.
\end{equation}

This follows by bounding the approximation residual under both the continuous measure and its discrete counterpart. The result explains why low-dimensional pricing sensitivities can lead to accurate lifts with few states, as investigated in RQ1 and RQ2.

These guarantees concern the first-order pricing problem. A small numerical rank alone does not establish a uniform approximation bound, and the existence of a positive quadrature does not guarantee that the nonlinear optimisation finds it. Moreover, the stochastic admissibility guarantees for positive exponential kernels do not automatically extend to signed weights. Our experiments therefore complement the theory by testing whether this low-dimensional structure can be exploited in practice.

\end{document}